\documentclass[onecolumn]{aa}
\usepackage{graphicx}
\usepackage{eufrak}
\usepackage{txfonts}
\usepackage{eufrak}
\usepackage{natbib}
\bibpunct{(}{)}{;}{a}{}{,}

\begin{document}

\title{Dust driven mass loss from carbon stars\\ as a function of stellar parameters}
\subtitle{I. A grid of Solar-metallicity wind models\\ \textit{(Corrigendum)}}
\titlerunning{Dust driven mass loss from carbon stars as a function of stellar parameters}

\author{Lars Mattsson\inst{1,2}\thanks{\email{mattsson@dark-cosmology.dk}}\and Rurik Wahlin\inst{2} \and Susanne H\"ofner\inst{2}}
\institute{Dark Cosmology Centre, Niels Bohr Institute, University of Copenhagen, Juliane Maries Vej 30, DK-2100, Copenhagen \O, Denmark
\and
Dept. of Physics and Astronomy, Div. of Astronomy and Space Physics, Uppsala University, Box 515, SE-751 20 Uppsala, Sweden}

\date{Received date; accepted date}
\abstract
{}
{The purpose of this corrigendum is to point out that a handful of models in the original paper (Mattsson, Wahlin \& H\"ofner 2010) were computed with faulty initial structures.}
{Using exactly the same modelling methods we have recomputed the faulty models with new initial structures.}
{The new results slightly changes some of the trends in the wind properties with stellar parameters, but the overall effects are small. The conclusions are not affected}
{}

\maketitle

While recomputing parts of the grid of wind models (Mattsson, Wahlin \& H\"ofner 2010) in order to obtain extended time sequences for modeling of spectral variation (Eriksson et al., 2012, work in progress) we discovered that a 
few models had faulty starting structures. A closer analysis revealed there were in total nine models that needed revision. We recomputed the faulty hydrostatic start models as well as the wind models and below we present a 
table (Table \ref{parameters}) with the resultant wind properties. 
Qualitatively, these new numbers have no major impact on the overall result, although they would slightly affect the appearance of Figs. 6 and 7.  The 
conclusions are not affected.

  \begin{table*}
  \caption{\label{parameters} Input parameters ($\Delta v_{\rm p}$, $M_\star$, $\log(L_\star)$, $L_\star$, $T_{\rm eff}$, log(C-O), $P$) and the 
  resulting average mass loss rate, average wind speed and the mean degree of dust condensation at the outer boundary. 
  The dust-to-gas mass ratio $\rho_{\rm d}/\rho_{\rm g}$ is calculated from $f_{\rm c}$ as described in H\"ofner \& Dorfi (1997). In the two cases where no numbers are given, no sustained outflows were produced according to the 
  model.}
  \center
  \begin{tabular}{lcccccccccccc}  
  \hline
  \hline

  $\Delta v_{\rm p}$ & $M_\star$ & $\log(L_\star)$ & $T_\mathrm{eff}$  & log(C-O)+12 & $P$ & $\langle\dot{M}\rangle$ & $\langle u_{\rm out} \rangle$ & $\langle {\rm f_c} \rangle$ & 
  $\langle {\rho_{\rm d}/\rho_{\rm g}} \rangle$\\[1mm]
  [km  s$^{-1}$] & [$M_\odot$] & [$L_{\odot}$] &  [$\mbox{K}$] & & [days] & [$M_\odot$ yr$^{-1}$] & [km  s$^{-1}$] & & \\

  \hline
\\
2.0 & 0.75 & 3.85 & 2600 & 8.80 & 393 & 4.96E-06 & 23.8  & 0.259 & 1.40E-03 \\
4.0 & 0.75 & 3.85 & 2600 & 8.80 & 393 & 6.54E-06 & 23.8  & 0.289 & 1.56E-03\\
6.0 & 0.75 & 3.85 & 2600 & 8.80 & 393 & 8.52E-06 & 23.6  & 0.321 & 1.74E-03\\[3mm]
2.0 & 1.00 & 4.00 & 2800 & 8.20 & 524 & -      &  -   &  - &  -\\
4.0 & 1.00 & 4.00 & 2800 & 8.20 & 524 & 6.74E-07 & 1.87  & 0.185 & 2.51E-04\\
6.0 & 1.00 & 4.00 & 2800 & 8.20 & 524 & 2.36E-06 & 1.69  & 0.204 & 2.77E-04\\[3mm]
2.0 & 1.00 & 3.85 & 3000 & 8.80 & 393 & -     &   -   &  -&  -\\
4.0 & 1.00 & 3.85 & 3000 & 8.80 & 393 & 1.43E-06 & 32.4  & 0.400 & 2.16E-03\\
6.0 & 1.00 & 3.85 & 3000 & 8.80 & 393 & 4.35E-06 & 20.0  & 0.293 & 1.58E-03\\
\\
  \hline
\\
  \end{tabular}
  \label{models}
  \end{table*}
  
\bibliographystyle{aa}

\end{document}